\documentclass{vgtc}                          % final (conference style)
\ifpdf%                                % if we use pdflatex
  \pdfoutput=1\relax                   % create PDFs from pdfLaTeX
  \usepackage{graphicx}                % allow us to embed graphics files
  \DeclareGraphicsExtensions{.pdf,.png,.jpg,.jpeg} % for pdflatex we expect .pdf, .png, or .jpg files
\else%                                 % else we use pure latex
  \usepackage{graphicx}                % allow us to embed graphics files
  \DeclareGraphicsExtensions{.eps}     % for pure latex we expect eps files
\fi%

\graphicspath{{figures/}{pictures/}{images/}{./}} % where to search for the images

\usepackage{microtype}                 % use micro-typography (slightly more compact, better to read)
\PassOptionsToPackage{warn}{textcomp}  % to address font issues with \textrightarrow
\usepackage{textcomp}                  % use better special symbols
\usepackage{mathptmx}                  % use matching math font
\usepackage{times}                     % we use Times as the main font
\usepackage{cite}                      % needed to automatically sort the references
\usepackage{tabu}                      % only used for the table example
\usepackage{booktabs}                  % only used for the table example
\onlineid{1171}

\vgtccategory{Research}

\vgtcinsertpkg

\title{Histogram binning revisited with a focus on human perception}

\author{Raphael Sahann\thanks{e-mail: raphael.sahann@univie.ac.at}\\ %
    \parbox{1.5in}{\scriptsize \centering Faculty of Computer Science, University of Vienna, Austria} %
\and Torsten M\"{o}ller\thanks{e-mail: torsten.moeller@univie.ac.at}\\ %
    \parbox{1.5in}{\scriptsize \centering Faculty of Computer Science, Data Science @ Uni Vienna, University of Vienna, Austria} %
\and Johanna Schmidt\thanks{e-mail: johanna.schmidt@vrvis.at}\\ %
    \parbox{1.5in}{\scriptsize \centering VRVis Zentrum für Virtual Reality und Visualisierung Forschungs-GmbH, Austria}}

\abstract{This paper presents a quantitative user study to evaluate how well users can visually perceive the underlying data distribution from a histogram representation. We used different sample and bin sizes and four different distributions (uniform, normal, bimodal, and gamma). The study results confirm that, in general, more bins correlate with fewer errors by the viewers. However, upon a certain number of bins, the error rate cannot be improved by adding more bins. By comparing our study results with the outcomes of existing mathematical models for histogram binning (e.g., Sturges’ formula, Scott’s normal reference rule, the Rice Rule, or Freedman–Diaconis’ choice), we can see that most of them overestimate the number of bins necessary to make the distribution visible to a human viewer.%
} % end of abstract

\keywords{empirical studies in visualization; histogram binning}

\teaser{
    \vspace{3mm}
    \includegraphics[width=\linewidth]{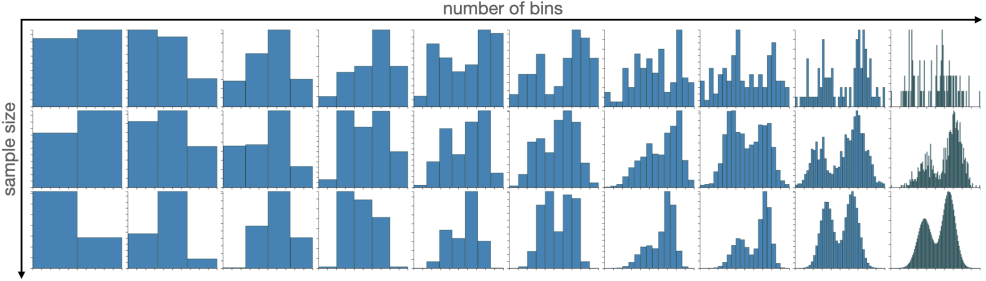}
    \centering
    \caption{Selection of histogram datasets used in our study. We evaluated how well human viewers can detect the underlying data distribution in a histogram when different sample sizes and bins are used. For this, we created datasets with a different number of samples (first row: few, last row: many) and a different number of bins (left column: $2$, right column: $100$). A bimodal distribution was used to create the datasets in this illustration.}
    \label{fig:teaser}
}

\begin{document}


%% \section{Introduction} %for journal use above \firstsection{..} instead
\firstsection{Introduction}

\maketitle

Histograms are a well-known and prevalent visualization technique~\cite{Nuzzo19} representing the distribution of univariate data by visualizing the tabulated frequency at certain intervals, represented as bars or \emph{bins}. Several bins next to each other help human viewers to build a mental model of the data distribution. The most important parameter visualization designers have to set when creating a histogram is the \emph{number of bins}.

Statisticians have developed several thumb rules to help researchers estimate the right number of bins when creating a histogram. For example, Sturge's formula~\cite{Sturges26} defines how to split the data into $k$ bins based on the number of samples being available. Scott's normal reference rule~\cite{Scott79} measures the discrepancy between the bin representation and the data distribution by employing mean integrated squared error. The Freedman–Diaconis choice~\cite{FreedmanD81} is based on minimizing the difference between the area under the data distribution and the area under the probability distribution defined by the binning. The so-called Rice’s rule~\cite{TerrellS85} can be applied for non-normal distributed data. More advanced approaches, like the ones by Lolla and Hoberock~\cite{LollaH11} and Birge and Rozenholc~\cite{BirgeR06}, use mathematical concepts like Cumulative Distribution Functions (CDF) or penalty functions to calculate the number of bins, reflecting the idea that smooth distributions need fewer bins than rough distributions.

The mathematical models, on the forefront being Sturge's formula, Scott's normal reference rule, the Rice’s rule, and the Freedman–Diaconis choice, are quite popular and are often used in current visualization systems and libraries. The mathematical models are fast and convenient binning estimations. Interestingly, the models and their suggested binnings have not been evaluated in perceptual user studies yet. The models in use today were statistically and mathematically evaluated. However, it is unclear how well the suggested numbers of bins match the human visual perception. The number of bins suggested by the mathematical models may be quite high ($> 300$). In case histograms need to be shown on small displays, e.g., smartphones, it would be interesting to know whether such a high number of bins is really needed when shown to a human viewer.

User studies~\cite{KosaraHIWL03} offer a scientifically sound method to measure how people read visualizations~\cite{Huang14}, and a number of studies have already been undertaken in an effort to assess these aspects~\cite{IsenIJSM13}. When it comes to summary statistics, Lem et al.~\cite{LemOVV13} and Kaplan et al.~\cite{KaplanGCM14} noticed a general problem for students when trying to read and interpret aggregated information in histograms and box plots. Correll et al.~\cite{CorrellLKS19} highlighted the importance of selecting the right number of bins for detecting missing values and outliers in a histogram. In general, the literacy of humans interpreting these visualizations seems to be decoupled from the statistical interpretation. According to Boels et al.~\cite{BoelsBVD19}, information reduction still seems to be an understudied topic in data visualization. This is also in line with the current need in research to understand how viewers can construct and interpret data visualizations~\cite{BoernerBG19} and with the need for further research on visualization guidelines~\cite{DiehlAEC18}.

% ---------------------------------------------------------------
\section{Quantitative User Study}

In this paper we report about a user study, also sometimes called user evaluation study~\cite{Forsell10}, that has been conducted to compare the number of bins suggested by mathematical models with human perception when analyzing histograms. We address the research question whether the numbers of bins suggested by statistical computations match the minimum number of bins required for human viewers to be able to detect the data's underlying distribution in a histogram. For this, we used datasets with four different distributions (\emph{uniform}, \emph{normal}, \emph{gamma}, and \emph{bimodal}), different sample sizes, and numbers of bins (see Figure~\ref{fig:teaser}), and asked participants to state which distributions they see in the different representations.

%-----------------
\subsection{Hypotheses generation}

%-------
\textbf{Task definition}: Histograms as summary statistics provide the possibility to perform several tasks related to distribution analysis (e.g., identifying the mean and the median or comparing quartiles). One task related to distribution analysis is to identify the data's underlying distribution, which has been classified as the task to ''\emph{describe and identify the shape and type of one distribution}'' in the literature~\cite{BlumeDM20}. The identification of the underlying distribution is the task we evaluated in our study.

%-------
\textbf{Distributions}: To get an overview of the distributions currently used in practice, we looked into literature targeted towards data scientists to learn more about the use of data distributions. Some examples are: \emph{Doing Data Science}~\cite{SchN13} lists $17$ density functions that data scientists should be familiar with. The \emph{Data Scientist’s Crib Sheet}~\cite{Owen18} describes $15$ density functions that are important and highlights their relationships. In the \emph{KDnuggets} tutorials~\cite{Strika19} five density functions are explained that data scientists should be aware of. Based on this literature research and based on our own experience when working with data, we decided to classify the available density distributions based on their main shape characteristics. We defined four main classes:
\vspace{-1mm}
\begin{itemize}
    \item \emph{uniform}: uniform distributions
    \vspace{-1mm}
    \item \emph{unimodal}: distributions with one peak, similar to a Gaussian kernel
    \vspace{-1mm}
    \item \emph{bimodal}: distributions with two peaks
    \vspace{-1mm}
    \item \emph{skewed}: distributions with one peak which are skewed to one side of the distribution
    \vspace{-1mm}
\end{itemize}

This classification is also confirmed by Walker~\cite{MathBootCamps17}, who describes the most common shapes of distributions as \emph{bell shaped}, \emph{left skewed}, \emph{right skewed}, \emph{bimodal}, and \emph{uniform}. A quick pre-test among students in the study preparation phase did not show any differences between the detection of left and right skewness. We, therefore, only considered distributions which are skewed to the left in our study. After deciding on the four classes, we identified one mathematical density function representing each class best:
\vspace{-1mm}
\begin{itemize}
    \item For the class \emph{uniform}, a uniform distribution fits best.
    \vspace{-1mm}
    \item For the class \emph{unimodal}, we selected the normal density function to represent this class.
    \vspace{-1mm}
    \item For the class \emph{bimodal} we joined two normal density functions to form a bimodal distribution with two peaks.
    \vspace{-1mm}
    \item For the class \emph{skewed} we selected the gamma density function to represent this class.
\end{itemize}

%-------
\begin{figure}[t!]
  \centering
  \includegraphics[width=\columnwidth]{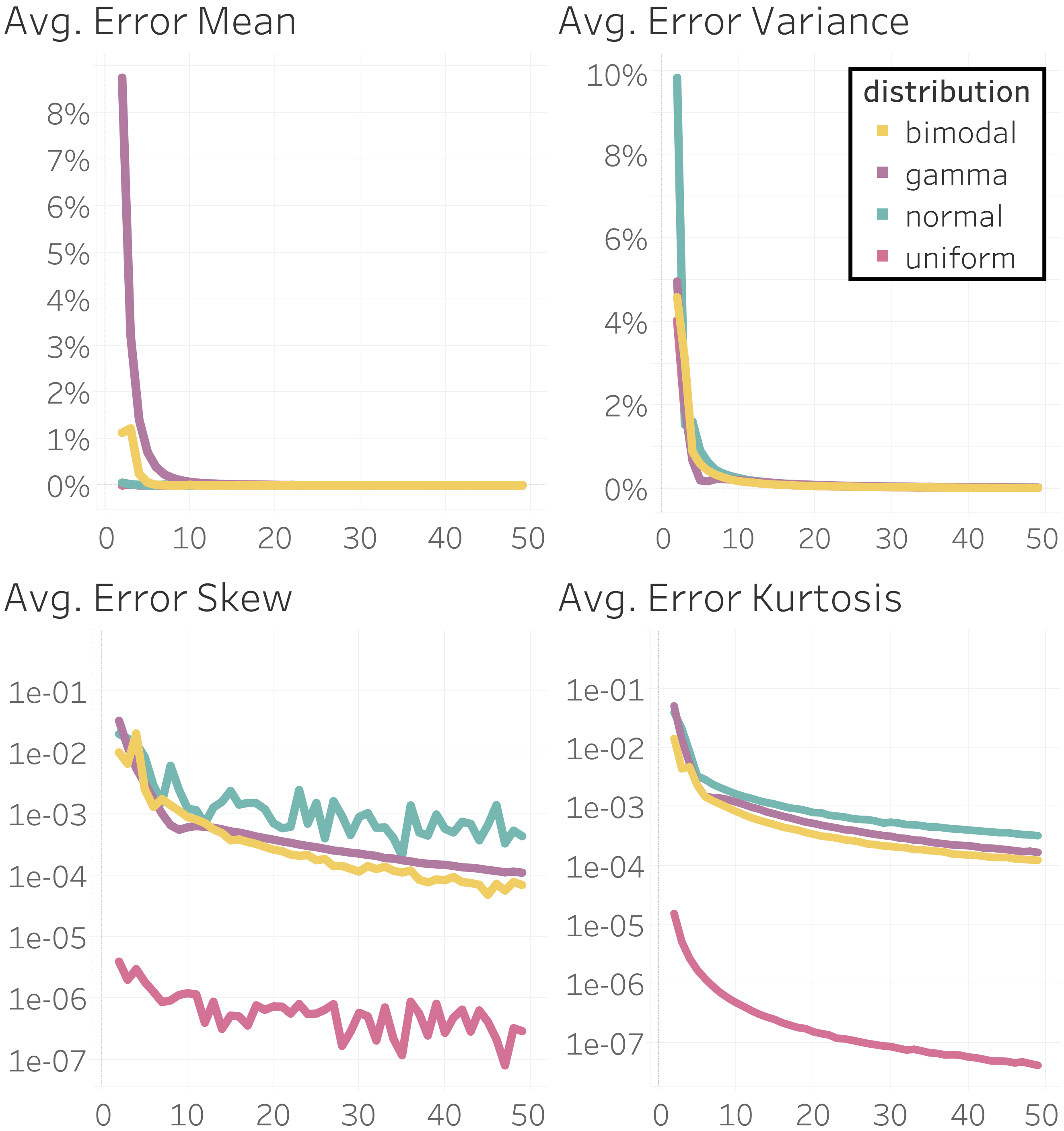}
  \caption{Calculated average errors of the four moments based on $100,000$ draws from binned samples ($n=1,000,000$) compared to the actual moments of the underlying distribution. The x-axis shows the number of bins, the y-axis shows the error in percent. Note that the y-axes are scaled differently.}  
  \label{fig:moments}
\end{figure}

%-------
\textbf{Number of samples and bins}: The number of samples and bins to be tested were chosen based on a mathematical analysis of the distributions. We calculated the four moments (mean, variance, skew, and kurtosis) for each of the distributions. The combination of these four moments can uniquely identify a distribution's shape. For five different sample sizes ($100, 1000, 10000, 100000$, and $1000000$) we drew $1,000$ times from the four distributions specified above and recorded the actual moments from each draw. We then created evenly spaced binnings in steps of $1$ from $2$ up to $100$ bins for each draw. Using the bins' centers as the outline of a new shape, we calculated its moments and compared them to the actual moments. Figure~\ref{fig:moments} shows the calculated errors in the four moments when using different bin sizes. The error is large for small numbers of bins for all moments but reaches an almost constant rate above $10$ bins. Adding more bins does not affect the error rate anymore. We, therefore, decided that the range up to $10$ bins would be of the greatest interest for our study. Since some moments (especially kurtosis) take a little longer to settle completely, we chose to include some values between $10$ and $40$ as well. We also included $100$ bins as an upper boundary since it is equal to our study's smallest sample size.

%-------
\textbf{Hypotheses generation}: We agreed upon testing the following hypotheses:
\begin{itemize}
    \vspace{-1mm}
    \item \textbf{Hypothesis H1}: The number of bins influences how well humans can perceive the underlying data distribution in a histogram.
    \vspace{-1mm}
    \item \textbf{Hypothesis H2}: Upon a certain number of bins, adding new bins does not improve the perception of underlying data distributions in a histogram.
\end{itemize}

%-----------------
\subsection{Study design}

To summarize, we tested
\vspace{-2mm}
\begin{itemize}
    \item four distributions (uniform, normal, bimodal, and gamma), with
    \vspace{-2mm}
    \item ten different bin counts ($2, 3, 4, 5, 7, 10, 15, 20, 40$, and $100$), and
    \vspace{-2mm}
    \item four sample sizes ($100, 1000, 10000$, and $1000000$).
    \vspace{-2mm}
\end{itemize}

We used the approach of a web-based questionnaire to be able to reach a large group of participants~\cite{Reips07}. A Cross-Site Request Forgery (CSRF) token was generated whenever a participant decided to start the survey. Since we then used only this token to identify the participant, the study was fully anonymized without any possibility to track the results back to the participants.

For every participant, we started with an initial explanation what the study will be about. Afterwards we included a sanity check to filter out careless participants~\cite{NiessenMT16}. Afterwards the actual study questions started, where we showed $20$ histograms to every participant, one after the other. The histograms were randomly selected from the pool of datasets. The histogram plots had two axes with ticks, but we did not show any numbers or scales. Participants were asked to answer the question
    
``\emph{1. Choose the distribution which resembles the image above most closely}``
    
by clicking on one of the icons below the histogram showing different possible distributions. Participants were also asked to state
    
\begin{figure}[t!]
  \centering
  \includegraphics[width=0.9\columnwidth]{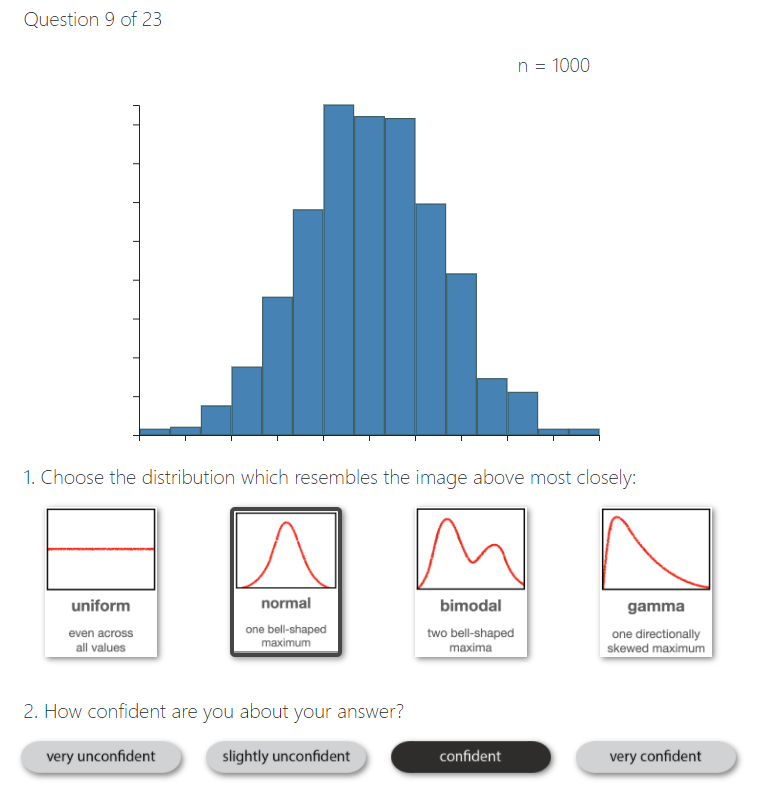}
  \caption{Study question. Participants were shown a histogram depicting the data's underlying distribution and asked to click the appropriate icon. Participants were also asked to state how confident they are about their answers. In this example \emph{normal} and \emph{confident} have been selected.}
  \label{fig:surveyquestion}
\end{figure}

``\emph{2. How confident are you about your answer?}``
    
on a four-point Likert scale. Participants could only proceed with the next histogram if they answered both questions. An example of how the web-based implementation of such a histogram question looked like is shown in Figure~\ref{fig:surveyquestion}. The study setting can best be described as a \emph{judgement study}, where the study's purpose is to gather a person’s response to a set of stimuli~\cite{Carpendale08}. According to the literature, judgement studies are a commonly used approach for perceptual studies.

%-------

% ---------------------------------------------------------------
\section{Results}
\label{sec:results}

In total, $82$ participants finished the user study within a $14$ day time frame. We only counted complete submissions and did not record the dropouts. Based on the sanity check questions, we had to exclude $10$ data records from the evaluation, which led to a final number of $72$ valid submissions.

The majority of the participants were between $20$ and $49$ years old. One-third of the participants ($33\%$) were bachelor's, master's, or PhD students. The other participants were working part-time ($14\%$), full-time ($23\%$), or, more specifically, in research and education ($26\%$). $27\%$ of the participants had some experience in reading charts and plots in the media. $27\%$ classified themselves as being experienced in reading data visualizations, and $41\%$ stated that they are also creating data visualizations themselves. Only four participants stated that they do not have any experience with data visualization.

%--------------------------------------
\subsection{Study results}
\label{sec:results:study}

The evaluation of the quantitative results led to the following results:

\vspace{1mm}
\textbf{Insight 1: Small sample sizes generally make it harder to detect the underlying data distribution, which can only slightly be mitigated by using a higher number of bins}.

For datasets with $100$ samples, $35.4\%$ of the answers were wrong. With $1000$, $10000$, and $1000000$ samples being available, the detection error rate could be halved to $16.1\%$, $18.3\%$, and $18.8\%$. A Mann–Whitney U test~\cite{FieldH03} resulted in p-values $p < 0.001$ when comparing the results for all sample sizes, which confirms the statistical significance of the results. Participants stated to be less confident when judging the distribution with a sample size of $100$. The amount of participants being \emph{very confident} about their answers constantly increases with a rate of about $10\%$ for larger samples sizes.

%-------
\begin{figure}[b!]
  \centering
  \includegraphics[width=\columnwidth]{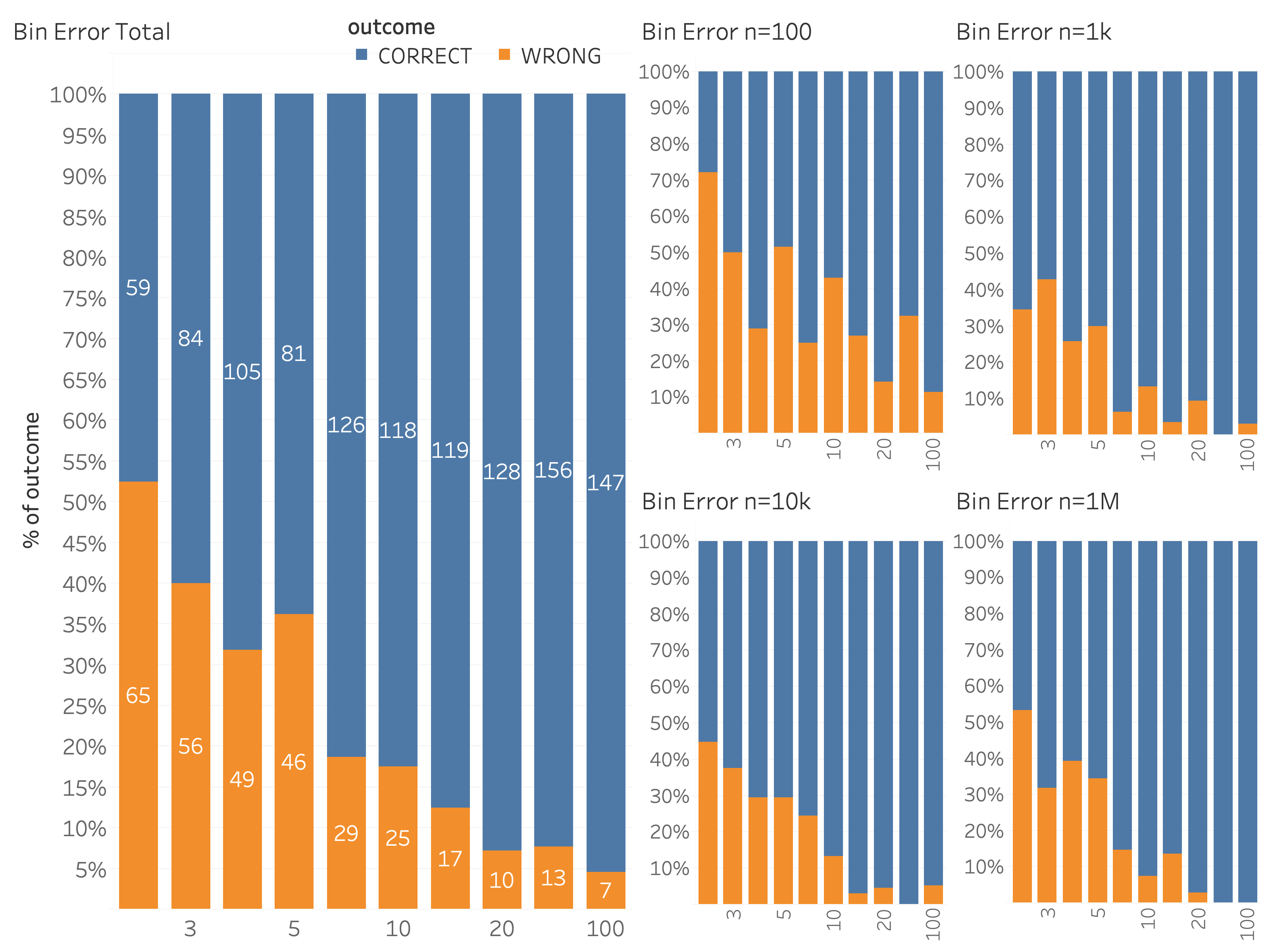}
  \caption{Correct (\emph{blue}) and wrong (\emph{orange}) answers based on the number of bins and sample size. With only $100$ samples, the data distribution recognition is generally challenging. For larger sample sizes, a larger number of bins increases the recognition of the correct distribution. However, beyond $20$ bins, the detection rate does not increase significantly anymore.}
  \label{fig:results01}
\end{figure}

\vspace{1mm}
\textbf{Insight 2: Beyond a certain number of bins the error rate stays constant and is not improved by adding more bins.}

Like the sample size, the number of bins affected participants' ability to recognize the underlying distribution correctly. More bins result in fewer errors being made by the participants (see Figure~\ref{fig:results01}). This effect is different, depending on how many samples are available. In the case of $100$ samples, the error rate stays rather high, also in cases where a higher number of bins was used. For other sample sizes, the error rate decreases in case a larger number of bins is used. However, for larger sample sizes, it can be seen that more bins do not improve the visual perception of the underlying data distribution. While the error rate is significantly better when comparing the bin size $2$ to other parameters ($p < 0.001$), the difference between a larger number of bins is not significant any more (\emph{bins/bins}: \emph{p-value} -- $15$/$20$: $p = 0.072$, $20$/$40$: $p = 0.442$, $40$/$100$: $p = 0.121$).

%-------
\begin{figure}[t!]
  \centering
  \includegraphics[width=\columnwidth]{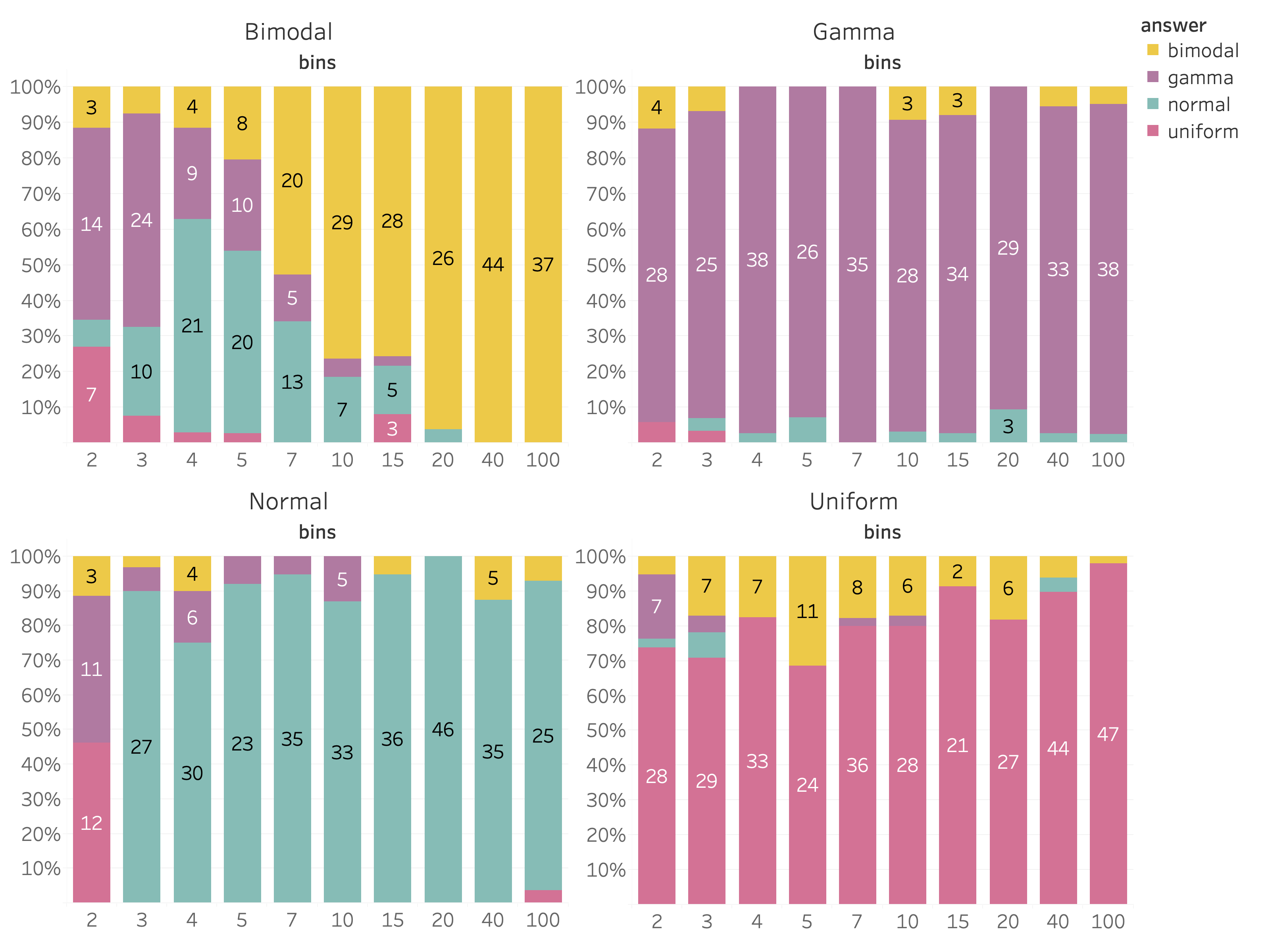}
  \caption{Recorded Answers by distribution and bin size. The chart title shows the actual shown distribution, the colored segments show the user responses. Users could mostly identify gamma and uniform distributions with two bins, needed three bins to distinguish a normal distribution, and they needed at least ten bins to recognize a bimodal distribution.}
  \label{fig:results02}
\end{figure}

\textbf{Insight 3: For bimodal distributions the number of bins is more important to recognize them correctly.}

The percentages of correct and wrong answers for every distribution indicate that it is generally easier to detect gamma and uniform distributions (see Figure~\ref{fig:results01}). The detection of bimodal distributions was harder and very strongly affected by the number of bins. In the case of normal distributions, apart from $2$ bins, the detection worked quite well. Bimodal distributions are often confused with gamma distributions in case of a low number of bins (see Figure~\ref{fig:results02}).

\vspace{1mm}
\textbf{Insight 4: Experience in reading data visualization had no impact on the error rate. More experience led to higher confidence when answering the questions.}

Only minor, non-significant differences could be detected when analyzing the percentage of correct and wrong answers compared to the participants' stated experience with visualizations. Participants with no or mediocre experience were generally less confident when answering the questions than those who had extensive data visualization knowledge.

\vspace{1mm}
\textbf{Insight 5: Most mathematical models overestimate the number of bins needed for characterizing distributions.}

When comparing the suggested bin count of different mathematical models shown in Table~\ref{tab:rules} to the error rates from our study (as shown in Figure~\ref{fig:results01}) it is clear that, apart from Sturge's formula, all models have a tendency suggest too many bins starting from $1,000$ samples. Our results show that much fewer bins are needed for detecting distributions.

%-------
\begin{table}[t!]
    \centering
    \resizebox{\columnwidth}{!}{%
    \begin{tabular}{rrrrrr}
    \hline
    & & Scott's & & Freedman- & \\
    & Sturge's & normal & Rice’s & Diaconis & our \\
    samples & formula & reference &  rule & choice &  results \\ \hline 
    100 & 8 & 9 & 14 & 18 & 20 \\
    1,000 & 11 & 20 & 29 & 38 & 20 \\
    10,000 & 15 & 43 & 62 & 80 & 20 \\
    1,000,000 & 21 & 200 & 287 & 371 & 20 \\ \hline
    \end{tabular}
    }
    \vspace{0.4em}
    \caption{Suggested bin sizes from different mathematical models based on the number of samples in the data set. Apart from Sturge's formula almost all models overestimate the number of bins needed for $1,000$ samples and above. }
    \label{tab:rules}
\end{table}

%--------------------------------------
\subsection{Hypotheses testing}
\label{sec:results:hypotheses}

We can summarize the results of the study based on our two hypotheses in the following way:
\begin{itemize}
    \vspace{-1mm}
    \item \textbf{Hypothesis H1}: We can \emph{partially confirm} that the visual perception of the underlying data distribution depends on the number of bins. At least with larger sample sizes (enough samples to resemble the underlying distribution), the recognition becomes better when using more bins.
    \vspace{-1mm}
    \item \textbf{Hypothesis H2}: We can \emph{confirm} that upon a certain number of bins, adding new bins does not improve the perception of the underlying data distribution. After $20$ bins, the error rate cannot be decreased significantly by adding additional bins.
\end{itemize}

% ---------------------------------------------------------------
\section{Impact and discussion}

The task that was tested in the study was on ''\emph{describing and identifying the shape and type of one distribution}''~\cite{BlumeDM20}. We, therefore, would like to emphasize, that all results are solely to be interpreted for this specific task.

We also asked ourselves how the study results for this task can probably also be explained by sampling theory. Binning can be seen as a way to \emph{sample} the original distribution. As known from sampling theory, it is impossible to reconstruct the original function with too few samples. We, therefore, transferred the four density distributions we used (uniform, normal, bimodal, and gamma) into the frequency domain. None of these is band-limited, hence, there is no concrete Nyquist frequency. As a baseline we, therefore, used a representation of each density distribution with $1000000$ uniform samples. We then compared this baseline to representations of the same density distribution with less ($5, 10, 15, 20, 40$, and $100$) samples ($=$ bins). For a comparison we measured the deviation (i.e., error) between the baseline and the binned representation in the frequency domain. The representation with only five samples ($=$ bins) stands out to have the most significant error. From ten samples ($=$ bins) on, the error starts to converge to zero. $100$ samples ($=$ bins) already ensure an error to the original representation very close to zero. It is important to note that the difference between five and ten samples ($=$ bins) is much more significant than the difference between $40$ and $100$. The comparison of the representations in the frequency domain are, therefore, very similar to our study results.

A retrospective analysis of the study setting showed that, due to the fact the questions were randomly selected, an almost equal distribution of answers per question could be achieved (\textit{bins}: \textit{percentage of answers} -- $2$: $9.02\%$, $3$: $9.63\%$, $4$: $10.79\%$, $5$: $8.72\%$, $7$: $10.49\%$, $10$: $9.94\%$, $15$: $9.57\%$, $20$: $9.33\%$, $40$: $11.89\%$, $100$: $10.61\%$). As a drawback, participants mainly already had experience with data visualization. In the future, we would, therefore, like to access a broader range of users with a new study, and we would also like to test other tasks related to histograms.

% ---------------------------------------------------------------
\section{Conclusion}

We presented a quantitative evaluation and comparison of mathematically defined numbers of bins for histogram with human perception. The mathematical models (e.g., Scott's normal reference rule, the Rice Rule, Freedman-Diaconis' choice) mostly overestimate the number of bins necessary for a correct perception for human viewers. With around $20$ bins, the error rate for human viewers to detect the data's underlying distribution becomes stable and does not improve by adding more bins.

\acknowledgments{
We thank the anonymous reviewers for their helpful comments and suggestions which helped to improve the paper. VRVis is funded by BMK, BMDW, Styria, SFG, Tyrol and Vienna Business Agency in the scope of COMET - Competence Centers for Excellent Technologies (879730) which is managed by FFG.}

\bibliographystyle{abbrv-doi}

\bibliography{vis-short-paper}
\end{document}